# The topological origin of anomalous transport: Persistence of β in the face of varying correlation length.

Yaniv Edery

Faculty of Civil and Environmental Engineering, Technion, Haifa, Israel

## Abstract

Traditional concepts for flow in porous media assume that the heterogeneous distribution of hydraulic conductivity is the source for the contaminant temporal and spatial heavy tail, a process known as anomalous or non-Fickian transport; this anomalous transport behavior can be captured by the β parameter in the continues time random walk (CTRW) framework. This study shows that with the increase in spatial correlation length, between these heterogeneous distributions of hydraulic conductivities, the anomaly of the flow reduces, yet the β value is unchanged, suggesting that there is a topological component to the flow field, captured by the β. This finding is verified by an analysis on the flow field, showing that the changes in the conductivity values have little effect on the flow field morphology, which points to the topological component in the flow.

Introduction

The physics of dissolved chemical transport in fractured and porous media has been the subject of intense research for several decades. This transport is inherent in natural and artificial systems as chemical transport in soils and fractured rocks [*Bear*, 2013; *Haggerty et al.*, 2001; *Raveh-Rubin et al.*, 2015], oil recovery[*Edery et al.*, 2018; *Lenormand et al.*, 1983], filtration [*Tufenkji and Elimelech*, 2004], fuel cells [*Pharoah et al.*, 2006] and even percolation of coffee [*Fasano and Talamucci*, 2000], to count a few. The traditional modeling for this transport combines the transport with Fickian dispersion resulting in the advection-dispersion equation (ADE) [*Bear*, 2013]. However, the ADE is not sufficient in capturing this complex transport due to its ubiquities anomalous (or ''non-Fickian'') nature. Therefore, deterministic and stochastic models are required. As such, there is an extensive literature that addresses theoretical and numerical modeling approaches, backed-up with experimental evidence, for this anomalous transport [*B Berkowitz and Scher*, 1997; *B Berkowitz et al.*, 2006; *Ciriello et al.*, 2015; *Ciriello et al.*, 2013; *Cirpka and Kitanidis*, 2000; *Cushman and Ginn*, 1993; *Dagan and Neuman*, 2005; *Haggerty et al.*, 2000; *Kang et al.*, 2014; *Le Borgne et al.*, 2008; *Morales-Casique et al.*, 2006a; b; *Roubinet et al.*, 2013; *Sanchez-Vila and Carrera*, 2004; *Willmann et al.*, 2008].

The anomalous transport characteristic is the long temporal tail in the solute breakthrough curves (BTC), which is the contaminant concentration over a cross-sectional integral surface measure or a monitoring borehole in time [*B Berkowitz et al.*, 2000; *Bijeljic and Blunt*, 2006; *Edery et al.*, 2015; *Naftaly et al.*, 2015; *Salamon et al.*, 2007]. Alternatively, anomalous transport can be recognized as the deviation of the contaminant spatial plume from a Gaussian [*Y Berkowitz et al.*, 2013; *Kang et al.*, 2011; *Margolin and Berkowitz*, 2002]. Both manifestations of anomalous transport were proved to be associated with the multiscale heterogeneity of the porous media, ranging from the micron size pores to the Darcy scale of the hydraulic conductivity field (or K field) [*Bijeljic and Blunt*, 2006; *Di Donato et al.*, 2003; *Edery et al.*, 2014].

Various models explored this relation between the multiscale heterogeneity and anomalous transport: multi-rate mass transfer (MRMT), mobile-immobile model (MIM) and continuous-time random walk (CTRW) to count a few [*B Berkowitz et al.*, 2006; *Haggerty and Gorelick*, 1995; *Schumer et al.*, 2003]. In this study, the focus is on two well-established approaches; the first is a numerical simulation of flow through a varying K field and analysis of the emergent preferential flows (PF) [*Mariethoz et al.*, 2010; *Moreno and Tsang*, 1994; *Russian et al.*, 2016; *Sanchez-Vila et al.*, 2006]. The second is the CTRW framework with its flexible interpretation and prediction of laboratory and field-scale transport settings [*Ciriello et al.*, 2013].

The effectiveness of the CTRW formulation is in relating its governing parameters to physical and statistical characteristics in geological materials [*Bijeljic and Blunt*, 2006; *Edery et al.*, 2016]. Therefore, when defining a truncated power law (TPL) as the temporal probability density function (PDF) - w(t), in the CTRW framework, there are physical reasonings behind the TPL parameters, as shown in many studies [*Bijeljic et al.*, 2011; *Bromly and Hinz*, 2004; *Kosakowski*, 2004; *Kosakowski et al.*, 2001; *Mettier et al.*, 2006]. This temporal PDF-TPL, w(t), captures a range of physical aspects for flow in porous media, as they propagate from the pore scale to the Darcy scale [*Lester et al.*, 2014; *Quintard and Whitaker*, 1994].

At the Darcy scale, it is the heterogeneity of hydraulic conductivity distribution in the K field, which is the origin of the anomalous transport, as observed in the BTC and plume propagation [*Edery et al.*, 2014; *Levy and Berkowitz*, 2003]. Therefore, it is important to investigate the K field heterogeneity and link it to the anomalous nature of the flow [*Edery et al.*, 2014; *Kang et al.*, 2015; *Salamon et al.*, 2007]. A numerical investigation of the K field heterogeneity distribution is generally made by a layout of hydraulic conductivities, log-normally distributed (the PF model). The variance of this log-normal distribution establishes the domain heterogeneity or K field variance. This conductivity distribution, combined with a pressure difference from the domain inlet to outlet, provides the velocity flow field. A particle tracking (PT) simulation interrogates the flow field, while particles exiting the domain form the BTC [*Bianchi and Pedretti*, 2017; *Delay et al.*, 2005; *Hansen et al.*, 2018]. Knowing the K field and flow field a priori is not sufficient to determine the solute/particle PF's, and it is virtually impossible to know in advance the sampling of the domain lowest conductivities. Therefore, additional parameters as the Péclet number, connectivity, isotropy, and correlation length, that influence the transport should be considered [*Bruderer-Weng et al.*, 2004; *Dentz and Bolster*, 2010; *Edery et al.*, 2016; *Edery et al.*, 2014; *Moreno and Tsang*, 1994; *Sanchez-Vila et al.*, 2006].

In the context of anomalous transport, the correlation length smoothing nature, on the k field, is most influential on the velocity fields, thus altering the anomalous transport nature [*Filipovitch et al.*, 2016; *Le Borgne et al.*, 2008]. Although equations relating the correlation structure, of Eulerian or Lagrangian velocities, to statistical moments of K are available, e.g., through moment equations, the link between these two quantities is generally not explicit, as can be seen from this partial list of references [*Comolli and Dentz*, 2017; *Dentz and Bolster*, 2010; *Guadagnini and Neuman*, 1999; *Tartakovsky and Neuman*, 1998; *Tyukhova and Willmann*, 2016; *Tyukhova et al.*, 2015]. Therefore, understanding how the correlation length affects the Fickian to non-Fickian flow is a long-standing, imperative question [*Phillips and Wilson*, 1989; *Purvance and Andricevic*, 2000; *Rehfeldt et al.*, 1992].

In this work, we will show that an increase in the correlation length leads to the transition between non-Fickian to Fickian transport in a heterogeneous K field. The analysis will show how the correlation length increase, affects the K field, and, as a result, the PF's. More importantly, it will also show consistent features in the conductivity field and PF's as the correlation length increases and how the CTRW-TPL framework captures it. These consistent features will be related to the topology of the conductivity field and mainly to the morphology of PF's.

## Methods

To investigate the transition between Fickian to non-Fickian transport, as the correlation length changes, a set of numerical simulations is performed to produce a layout of 2D conductivity field distributed by a log-normal distribution with mean $\ln(k) \sim 0$ and variance $\sigma^2 = 5$, using a sequential Gaussian simulator (GCOSIM3D) [*Gómez-Hernández and Journel*, 1993]. This conductivity field is made of 120×300 conductivity bins (each with a size of $\Delta = 0.2$, therefore, the field size is 24×60), extracted from the middle of a 1200×3000 field to avoid boundary condition effects imposed by the correlation length function. Each field is produced by a statistical homogenous and isotropic Gaussian field in the $\ln(k)$, with a dimensionless correlation length l/L, produced by an exponential covariance, that ranges from $5 \times 10^{-3}$, weakly correlated, to $5 \times 10^{-2}$, strongly correlated, where L is the field length. As the aim is to investigate the effect of correlation length change in a consistent way, $\Delta/l$ ranges from 0.06 to 0.6. This range is on the high side for providing an accurate description of the small-scale fluctuations generated by the $\ln(K)$ field and advective transport [Ababou et al., 1989; Riva et al., 2009]. However, the study aims to investigate the transition from the uncorrelated or weakly correlated state to the highly correlated field. A complementary analysis with $\Delta = 0.02$, and therefore $\Delta/l < 0.1$ for all correlation lengths, is presented in the appendix and verifies the consistency of the results.

For each correlation length, 1000 realizations where produced, and a stability test was done to verify the mean conductivity. Each realization had a deterministic pressure drop ($\Delta P = 100$) imposed from the inlet (left) to the outlet (right). The local head difference for each conductivity bin is calculated using a finite element numerical model with Galerkin's weighting function [*Guadagnini and Neuman*, 1999]. Thus, the streamline for each realization is retrieved and from it, the local velocities, given that the porosity $\theta = 0.3$.

Each streamline realization is used to model solute transport by a standard Lagrangian, particle tracking model from which a break-through curve (BTC) is produced [*Bianchi et al.*, 2011; *Le Borgne et al.*, 2008]. For each realization, $10^5$ particles are flux-weighted at the inlet according to the inlet permeability distribution. The displacement size $\delta s$ is selected to be an order of magnitude less then $\Delta$, so to sample the velocity interpolation within a bin [*Cordes and Kinzelbach*, 1992]. At t=0, particles advances according to the local advection and diffusion term, by the following Langevin equation: $\mathbf{d} = \mathbf{v}[\mathbf{x}(t_k)]\delta t + \mathbf{d}_D$, where $\mathbf{d}$ is the displacement, $\mathbf{x}(t_k)$ is the particle known location at time $t_k$, $\mathbf{v}$ is the fluid velocity at that location, $\delta t = \delta s/v$ is the temporal displacement magnitude (v is the modulus of $\mathbf{v}$) and $\mathbf{d}_D$ is the diffusive displacement.

This diffusive displacement is randomly generated from a normal distribution between 0 to 1, multiplied by the square root of the diffusion coefficient ($D_m=10^{-5}$ cm$^2$/sec representing the diffusion of ions in water [*Domenico and Schwartz*] and the displacement magnitude as illustrated from the following equation: $\mathbf{d}_D=N[0,1]\sqrt{2D_m\delta t}$. The simulations were verified using $10^6$ particles and $\delta s < \Delta/10$ with no significant numerical dispersion.

The BTCs are produced for each simulation by registering the rate of particles at the outlet (after 300 unit length) in time for each realization of the heterogeneous domain. These BTCs represent a cross-sectional temporal distribution for particles arriving from various preferential flows. As such, a BTC from a single realization seems abrupt and not continues since particles are funneled towards a preferential flow, and from each preferential flow, the funneled particles exit at different times [*Edery et al.*, 2014]. To reach a continuous BTC flux of particles, the BTCs from the 1000 realizations are average. The averaged BTC is then normalized by the total number of particles, producing a normalized BTC. This process is reproduced for each correlation length in the study to allow comparable datasets.

The averaged BTC is then fitted with a CTRW-TPL equation based on a decoupled form of the transition time pdf, i.e., $\psi(\mathbf{s},t) = \psi(t)p(\mathbf{s})$. The transport equation in Laplace space (denoted by ~ and Laplace variable u) is $u\tilde{c}(\mathbf{s},u) - c_0(\mathbf{s}) = -\tilde{M}(u)[v_\psi \cdot \nabla\tilde{c}(\mathbf{s},u) - D_\psi : \nabla\nabla\tilde{c}(\mathbf{s},u)]$. Where $v_\psi = \frac{1}{\bar{t}}\int_0^\infty p(\mathbf{s})\mathbf{s}d\mathbf{s}$ is the particle transport velocity (distinct from the fluid velocity), $D_\psi = \frac{1}{2\bar{t}}\int_0^\infty p(\mathbf{s})\mathbf{s}^2 d\mathbf{s}$ is a generalized dispersion tensor in one dimension, $\tilde{M}(u) = \bar{t}u \frac{\tilde{\psi}(u)}{1-\tilde{\psi}(u)}$ is a "memory" function and $\bar{t}$ is a characteristic transition time [*B Berkowitz et al.*, 2006] and references therein). The function $\psi(t)$ is defined as the probability rate for a transition time t between sites, and as such, it determines the nature of the transport.

For this study, $\psi(t)$ is defined as a TPL form, written as $\psi(t) = \frac{n}{t_1}\frac{\exp(-t/t_2)}{(1+t/t_1)^{1+\beta}}$, $0 < \beta < 2$, where $n = [\left(\frac{t_1}{t_2}\right)^\beta e^{\frac{t_1}{t_2}}\Gamma(-\beta,\frac{t_1}{t_2})]^{-1}$ is a normalization factor, $\beta$ is a measure of the dispersive nature of the transport, $t_1$ is the characteristic transition time, $t_2$ is a "cut-off" time ($>t_1$) and $\Gamma(a,x)$ is the incomplete gamma function. Note that $\psi(t) \sim (t/t_1)^{-1-\beta}$ for $t_1 \ll t \ll t_2$, and decreases exponentially $\psi(t) \sim e^{-t/t_2}$, Fickian flow, for $t \gg t_2$; Fickian transport also occurs for $\beta > 2$.

## Results

### The correlation length effect on the conductivity field and BTC

To appreciate the effect of the correlation length has on transport, it is imperative to understand how the correlation length affects the conductivity field. Since the correlation length defines how well two bins are correlated in the conductivity domain, as a function of their distance, it is not surprising to see that increasing the correlation homogenizes the conductivity values locally, as can be seen in Figure 1a-c. Moreover, the increasing correlation homogenizes the features for the same field distribution, forming inclusions with similar permeabilities. Given this conductivity smoothing, due to the increase in the correlation function, the expectation is that the flow will become more Fickian in nature, and the tails of the BTC will become less pronounced. Yet in what way the transition between Fickian to non-Fickian transport takes place? Increasing the correlation length for the same log-normal conductivity fields must be reflected in one or several parameters in the CTRW-TPL adaptive distribution function.

To verify the CTRW-TPL adaptivity, the CTRW Matlab toolbox is used to fit the BTCs of various correlation lengths with the TPL-PDF. As can be seen from Figure 2a, indeed, the tailing of the BTC decreases with the increase in correlation length, echoing the transition to a more Fickian flow. However, the CTRW-TPL fits reproduce this transition not by the β, the parameter related to the heterogeneity, but by the $t_2$ that represents the cutoff of the power-law tailing and the transition to fickian flow. As the correlation length increases, β remains constant at 1.63, yet $t_2$ decreases logarithmically, portraying the fastest rate of change (see figure 6 in the appendix for a larger range of correlation lengths).

Not only the $t_2$ changes with the correlation length, the mean weighting time, $t_1$ also increases as the correlation length increase. Since the range between $t_1$ and $t_2$ defines the power-law tail (β) extant, the influence of β reduces with the correlation length without changing the actual value of β. The change in $t_1$ is correlated with the change of v and D by the first and second moment calculation; this correlation is apparent in Figure 2a BTCs. Specifically, as the correlation length increases, the BTC peak arrives at an earlier time while the BTC standard deviation increases. Therefore, the early peak arrival in figure 2a, is related to the increase in v, while the increase in the BTC standard deviation is an outcome of the increase in D. Thus, the β is unchanged while the correlation length increase, yet the range at which this β is relevant is decreasing due to the decreasing range of $t_1$ and $t_2$. Furthermore, the dispersion coefficient and velocity are increasing with the correlation length increase.

As verification for the CTRW parameter, the temporal pdf from the weighted conductivity bins and local head difference, as developed in Edery et al. 2014, is calculated for all correlation lengths. The weighting time for all realizations is integrated into one vector, with a corresponding vector for the frequency of the weighting time weighted by the particle visitation. A histogram of that weighting time vector was produced and normalized by the histogram bin size to produce the pdf of the realization ensemble. The $\psi(t)$ with the fitted parameters in figure 2a is juxta positioned, proving that the parameter are emergent from the PT simulation (see figure 2b-d for details). Moreover, the pdf slope, on a log-log scale, follows $-1-\beta$, this slope was similar for all the correlations with the value of $\beta = 1.63$.

The relation between correlation length to flow field

A particle visitation per bin investigates the effect of the correlation length increase on the flow field. The particle visitation entails a procedure where each particle visiting a bin is registered once, thus representing how many particles are visiting each bin throughout the simulation. Surprisingly, while the increase in correlation length smooths the transition of permeability between bins, fewer particles transition between bins. As such, the preferential flow is more distinct and tortures, as seen in figure 1d-f, the opposite from the expectedly uniform flow field. This implies that the increase in correlation length does not "spread out" the flow, as previously suggested, but rather funnels it and makes it more compact. Namely, for low correlation length, there are more preferential flows with fewer particles per bin on average. As the correlation length increase, the preferential flows are limited, compact, and hold more particles per preferential flow. There are also considerably more voids, places where there is zero particle visitation, and less uniformity of the number of particles per bin.

This spreading of the preferential flows is verified in a subsection of 100 realizations from the 1000 realizations made for each correlation length. The main apparent features for this process are the fact that the conductivities are not only smoothened as correlation length increases but that the edges of the conductivity decrease. They are forming a smoother transition for the particles between bins and therefore reducing the bifurcation of the flow. As such, it seems that as the correlation length increase, there is a smoother connected path for the particles. Therefore, the optimal preferential flows are dominant, and the number of preferential flows reduces. Decreasing the correlation length creates a more homogenous distribution of conductivities in the sense that there is less local dominance per permeability. Therefore, the number of preferential flows increase since they bifurcate frequently.

Another unexpected feature is that the preferential flows are persistent in their location regardless of the correlation length. As such, for conductivity fields that are generated by the same random seed yet smoothen by different correlation lengths, preferential flows appear roughly in the same location, although the conductivity they sample is different. This finding indicates that the relative arrangement is important for conductivities, and this arrangement is manifested, in a non-trivial way, by the preferential flow morphology.

Quantifying the correlation length effect on the flow field

To quantify the connection between the conductivity change, due to the correlation length, and the morphological change in flow field, a comparison between the conductivity distribution to the conductivity distribution weighted by particle visitations, is made for each correlation length. Weighting each conductivity distribution by the particle visitation provides the conductivity distribution as sampled by the flow, and in many ways, it is the effective conductivity distribution. Therefore, this comparison between the effective conductivity distribution to the original one is averaged over all 1000 realization, so to produce the mean-field conductivity distribution (see figure 3, a-c).

The variance of the log-normal conductivity field is slightly decreasing, as the correlation length increases, which is to be expected by the correlation length smoothing effect. However, it seems that the conductivity mean for the weighted and un-weighted conductivity distribution increases non-monotonically with the correlation length. Therefore, this non-monotonic increase in mean conductivity is linked to the increase in the correlation length smoothing effect for the un-weighted mean. As for the weighted mean, the correlation length increase leads to less preferential flow with more particles, as previously discussed. Therefore, a smaller subset of conductivities is sampled, and the mean for this conductivity subset is higher.

A careful look at the preferential flows and how they change in figure 1d-f shows something interesting; indeed, as the correlation length increases, the number of preferential flows decreases, yet there are preferential flows that are persistent in their location regardless of the correlation length. These persistent preferential flows (PPFs) that are shared for all correlation lengths are revealed by marking the cells visited by particles for the same realization field yet with varying correlation lengths simulation, as seen in figure 3 d-f. Focusing on these PPFs in figure 3 d-f, showed that these preferential flows are 40% of the entire domain, and between 50% to 80% of these PPF's are commune between the various correlation lengths. This means that the PPF is dominant for various correlation lengths and are persistent in their location.

The number of particles visiting these shared preferential flows is at the order of $10^3$-$10^4$ particles (or 1-10 % of all the particles), as can be seen in Figure 3 d-f, yet the unweighted permeability distribution on these PPFs is similar to the domain permeability distribution (not shown here). Probing the permeabilities within these PPFs weighted by the visiting particles shows that the PPFs low conductivities overlay on the total field weighted permeability, yet the PPFs high permeabilities overlay on the permeabilities of the total field un-weighted permeability distribution for all correlations, as can be seen by the green asterisk in figure 3 a-c. The mean and variance of the weighted permeabilities on the PPFs coincide with this finding, the variance is smaller while the mean is between the weighted and unweighted mean, and both are associated with the correlation length increase. Therefore, these preferential flow locations are common for all correlation length, not due to the absolute permeability value at their path, but rather by the permeability location and relative value, or in other words, by the topological arrangement of these permeabilities. This finding is verified for 100 realizations and proved to be the case for all of them. The shared preferential flows for the same realization seed with varying correlation length is 35%-45% from the whole domain available for flow and between 50% to 80% of the total preferential flows in total, while the absolute value of permeabilities changes between correlation lengths.

This topological relation between permeability is emergent by the flow and is not solely related to the absolute value of permeabilities. However, it should display a domain characteristic as portrayed by Euler characteristic [*Vogel*, 2002]. This Euler characteristic should be coupled with the flow and Péclet number and not necessarily with the permeability [*Nissan and Berkowitz*, 2019]. Incorporating the Euler characteristic with the analysis shown here is important and timely, yet it is out of the scope of this study; as such, it will be the subject of a future study.

The topological persistence of the power-law

The emerging question from this analysis is, what is the connection between the PPFs to the persistent β? Indeed, the β is inherently associated with the heterogeneity variance and, as such, for low correlation length, it reflects the range and variance of conductivities. However, as was shown in Figures 1 and 3, the range and variance of conductivities decrease with the increase in correlation length. Therefore, the same β is valid for various ranges of variance. The only feature that is consistent among the conductivities in figure 1a-c, as the correlation length increases, is the geometrical layout of conductivities and the relative value and location of conductivities in the 2D space, as portrayed by the PPFs. The same spatial layouts of conductivities are consistent for various correlation lengths; only their absolute values change, yet their ratio with adjacent bins is preserved.

The fact that the spatial geometry of the conductivities and preferential flow is preserved, as is β, while the correlation length varies, suggests that the power-law feature in the TPL distribution reflects the geometrical layout of conductivities. In a way, this must be the case since the preferential flows are sampling a subset of the conductivities in the domain, and this subset is mostly shared for all correlation lengths and is reflected by the β in the power law. Moreover, in a previous study, *Edery et al., 2015* showed that the relation between the β to the log-normal distribution variance, or heterogeneity, is within the bounds of the $t_1$-$t_2$ range, therefore, while the correlation length changes this range is changing but not the value of the β [*Edery et al.*, 2014]. This reduction in the $t_1$-$t_2$ range is apparent in the juxtaposition of the PDF in figure 2b-d. Hence the power law is consistent with the topology of conductivities in a statistical way, at the subset of the preferential flow.

Given that the β is unchanging and reflecting the unchanging topological layout of the conductivity subset, as correlation length changes, the $t_1$ - $t_2$ range reflects the range of subset conductivities yet not their distribution. This range change suggests that when a particle, within a preferential flow, starts experiencing the same conductivities at a constant rate, the particle reached the cutoff for the TPL tail. The smoothing of conductivities, achieved by the correlation length increase, suggests exactly that a particle within a preferential flow starts sampling the same range of conductivities in its subset. The smoothing of the conductivities also explains the un-intuitive increase in the dispersion (D) as the correlation length increase; the preferential flow spreads out for lower correlation lengths which would suggest more dispersion, yet, the dispersion in conductivity values is decreasing along within the preferential flow since the conductivity range is more uniformly distributed. Therefore, one should not confuse spatial spreading with dispersion in the CTRW framework, contradictory to the ADE framework.

Discussion

In this study, the role of the correlation length in a heterogeneous conductivity field is explored. An analysis of the variation in the conductivity fields and preferential flow is investigated both qualitatively and quantitatively. A CTRW-TPL framework is used to match the BTCs in the context of the conductivities and particle visitations. The work can be divided into two main parts, the spatial effects of the correlation length, the temporal effect captured by the CTRW-TPL, and the connection between them.

Starting with the spatial effect, it is clear by this study that the increase in correlation length not only smooths the conductivity in the domain but unintuitively funnels the flow to more distinct preferential flows. As the preferential flows become more distinct, the permeability values sampled by the flow changes, yet the geometrical layout of conductivities is unchanged and has the same spatial relation. The flow of particles, as manifested by the particle visitations scheme, spreads less as correlation length increases. Therefore, each preferential flow carries more particles, and the flow is less uniform. Even so, the preferential flow locations are persistent, suggesting that although their permeability value changes, the spatial permeability ratio is preserved. Quantifying these effects by the histograms shows that indeed the conductivity variance is effectively decreasing as correlation length increase, without changing the log-normal distribution variance, while the mean conductivity sampled by the preferential flow becomes closer to the log-normal distribution mean.

These spatial effects are captured by a decrease of the $t_1$-$t_2$ range in the CTRW-TPL, as the correlation length increase while the β value stays constant, regardless of the absolute change in conductivity variance. Concluding that the β is representing the topological layout of conductivities and not only their variance. The findings in this study unravel the complex effect of correlation length on the Fickian to non-Fickian flow [*Phillips and Wilson*, 1989; *Purvance and Andricevic*, 2000; *Rehfeldt et al.*, 1992] while tying the topological nature of the β with the Euler number for the preferential flows will be explored in a future study.

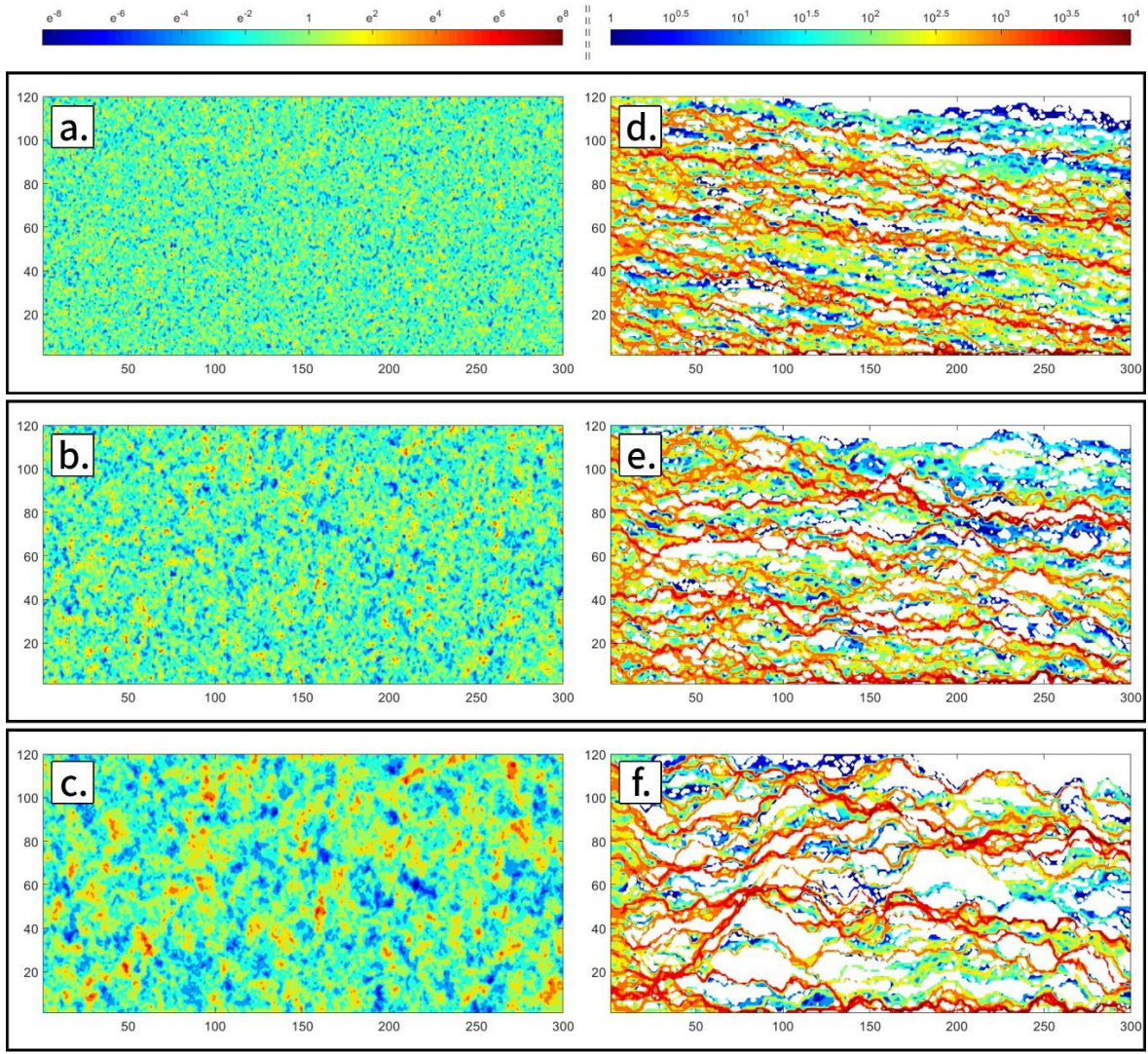

*Figure 1:* a-c) Three conductivity fields with varying correlation lengths of *5×10⁻³, 1.6×10⁻², and 5×10⁻²* , respectively generated with the same seed, presented on a log-normal scale. d-f) Particle visitation per bin for the a-c realizations, presented on a Log10 scale. As can be seen, while the conductivity field becomes more correlated and inclusions of conductivities increase in size and smoothen in values, the preferential flows, portrayed by the particle visitation, becomes sparser and less homogenous

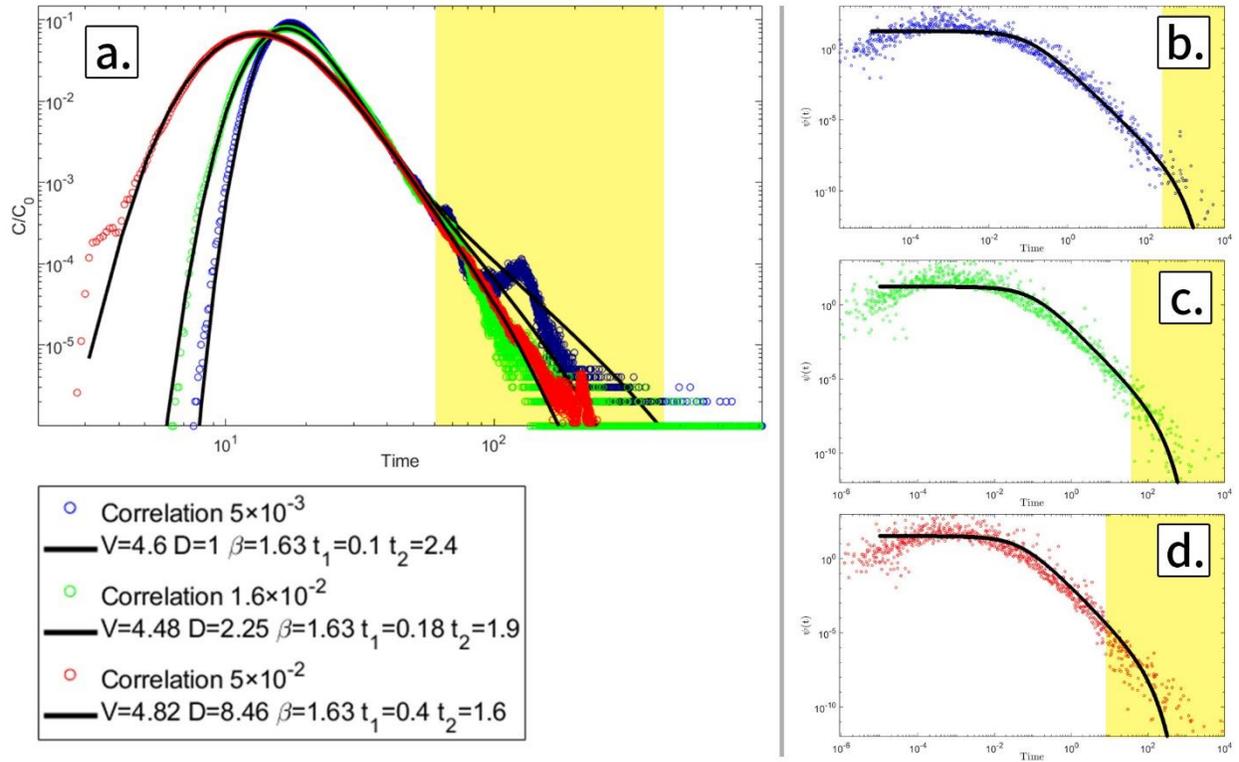

Figure 2: a) Ensemble BTCs for correlation lengths of $5\times10^{-3}$, $1.6\times10^{-2}$, and $5\times10^{-2}$ with CTRW-TPL fits for each. As can be seen, the tail decreases, yet it is the $t_2$ that displays the biggest changes. c-d) The PDF of $5\times10^{-3}$, $1.6\times10^{-2}$, and $5\times10^{-2}$, respectively, for the ensemble, weighted by the particle visitation with the juxtaposition of the CTRW PDF with the same parameters found for a-c. As can be seen, the logarithmic slope, which corresponds to $\beta$, is identical while the $t_2$ tailing, highlighted in yellow for the PDF and BTC figures, reduces with the increase in correlation length.

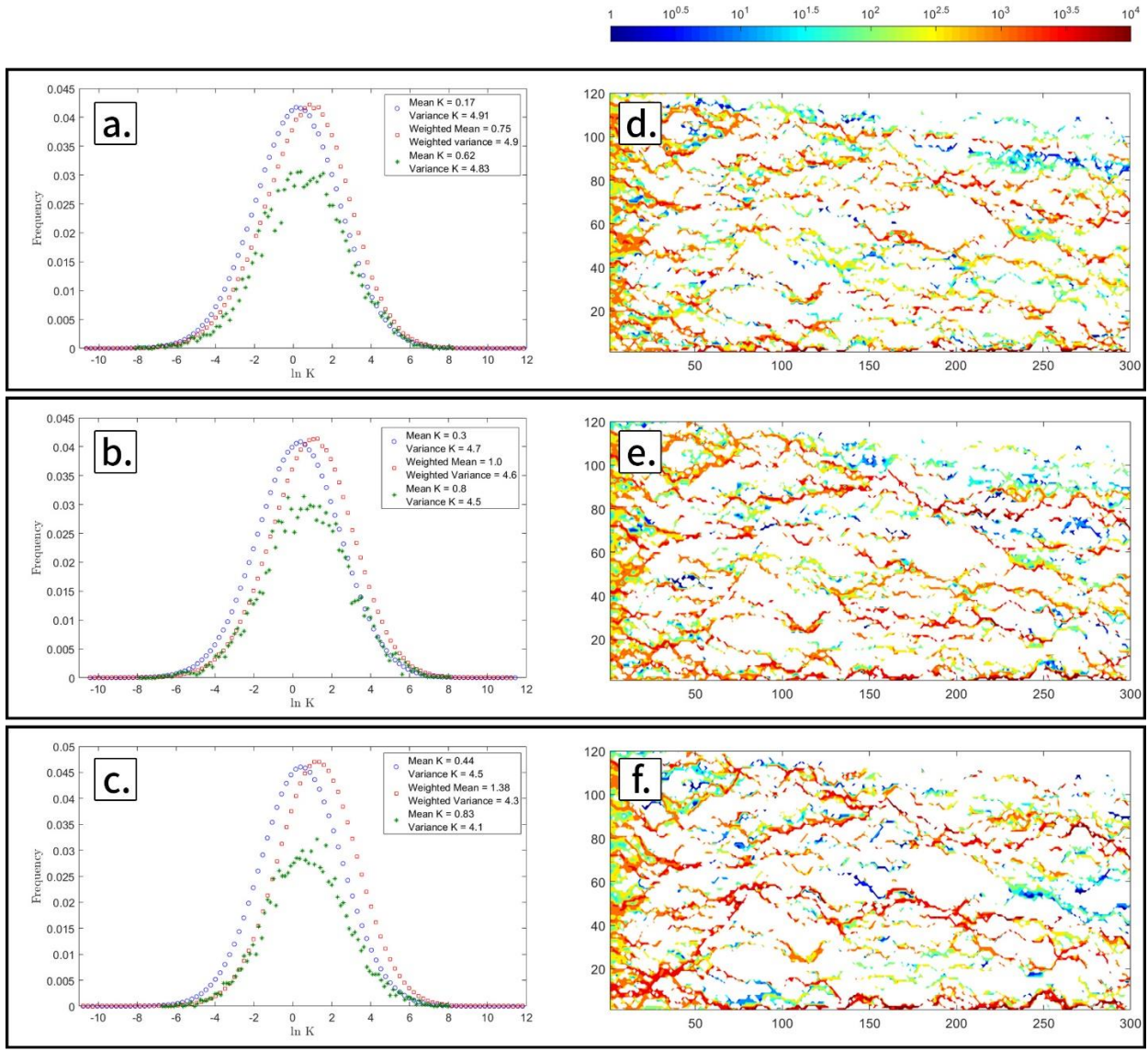

*Figure 3 a-c) Blue circles are the conductivity distribution for correlation lengths of 5×10$^{-3}$, 1.6×10$^{-2}$, and 5×10$^{-2}$ respectively, averaged for 1000 realizations. Red squares are the same conductivity distributions only weighted by the particle's visitations at each bin. Green asterisks are the permeabilities weighted by the particle visitations only on the PPF. d-f) Particle visitations on the PPF for correlation lengths 5×10$^{-3}$, 1.6×10$^{-2}$, and 5×10$^{-2}$ respectively.*

## **Appendix**

As stated in the method section, a parallel simulation is presented for $\Delta=0.02$. As previously, we use a field of 3000×1200 bins from which a 300×120 field is taken. Unlike the previous simulation, the absolute field size is now 6×1.2, which is an order of magnitude smaller. Thus, the head difference was reduced from 100 to 10 so to maintain the same mean velocity. As such, the dimensionless correlation length (normalized by the field length) ranges from $1\times10^{-5}$ to $1\times10^{-4}$, yet there are 15 bins for the low correlation length and 150 bins for the high correlation length.

As can be seen in by the BTC fit in figure 4a-c and the corresponding juxtaposition in figure 4d-f, the β is still persistent for the variations in correlation length while the $t_2$ is reducing due to the correlation length increase as before. Moreover, the same dispersion increase can be found with an increase in correlation length. This shows the consistency of these results, mainly since the juxtaposition in figures 4d-f still reproduces the same slope, which is associated with the β while the $t_2$ cutoff is apparent in the PDF tail dispersion. Repeating the same analysis for the PPF's as the correlation length changes shows even higher values of PPF's spatial occurrence for varying correlation length of more than 60% to 85% (see figure 5). As previously, the weighted conductivity distribution for the PPF's has the same tailing as in figure 3a-c, where the backward tail follows the weighted conductivity, and the forward tail follows the unweighted distribution (see figure 5a-c).

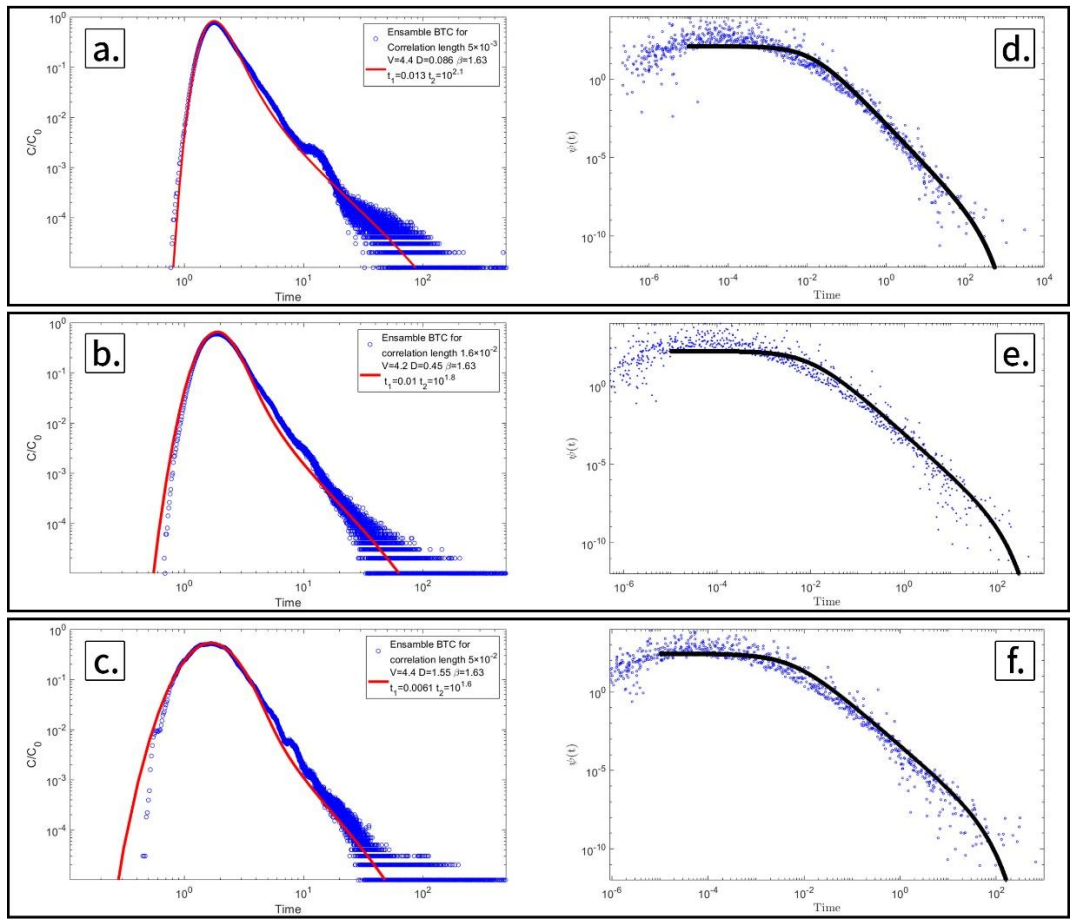

*Figure 4: a-c) Ensemble BTCs for correlation lengths of $5\times10^{-3}$, $1.6\times10^{-2}$, and $5\times10^{-2}$ with CTRW-TPL fits for each. As can be seen, the tail decreases, yet it is the $t_2$ and D that displays the biggest changes. d-f) The PDF of $5\times10^{-3}$, $1.6\times10^{-2}$, and $5\times10^{-2}$, respectively, for the ensemble, weighted by the particle visitation with the juxtaposition of the CTRW PDF with the same parameters found for a-c. As can be seen, the logarithmic slope, which corresponds to β, is identical while the $t_2$ tailing and the $t_1$ mean reduces with the increase in correlation length.*

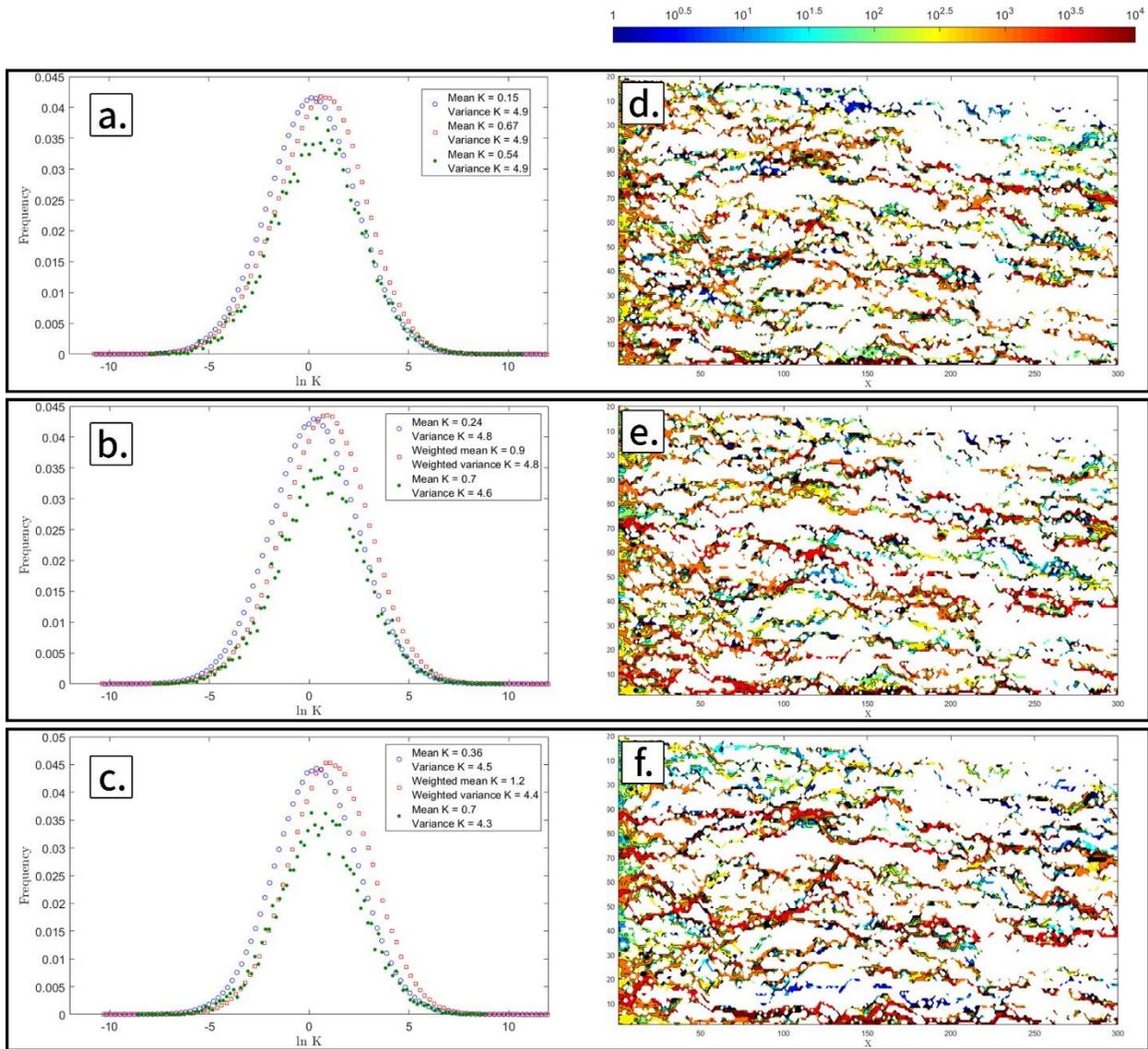

*Figure 5. a-c) Blue circles are the conductivity distribution for correlation lengths of 5×10⁻³, 1.6×10⁻², and 5×10⁻², respectively, averaged for 1000 realizations. Red squares are the same conductivity distributions only weighted by the particle's visitations at each bin. Green asterisks are the permeabilities weighted by the particle visitations only on the PPF. d-f) Particle visitations on the PPF for correlation lengths of 5×10⁻³, 1.6×10⁻², and 5×10⁻², respectively.*

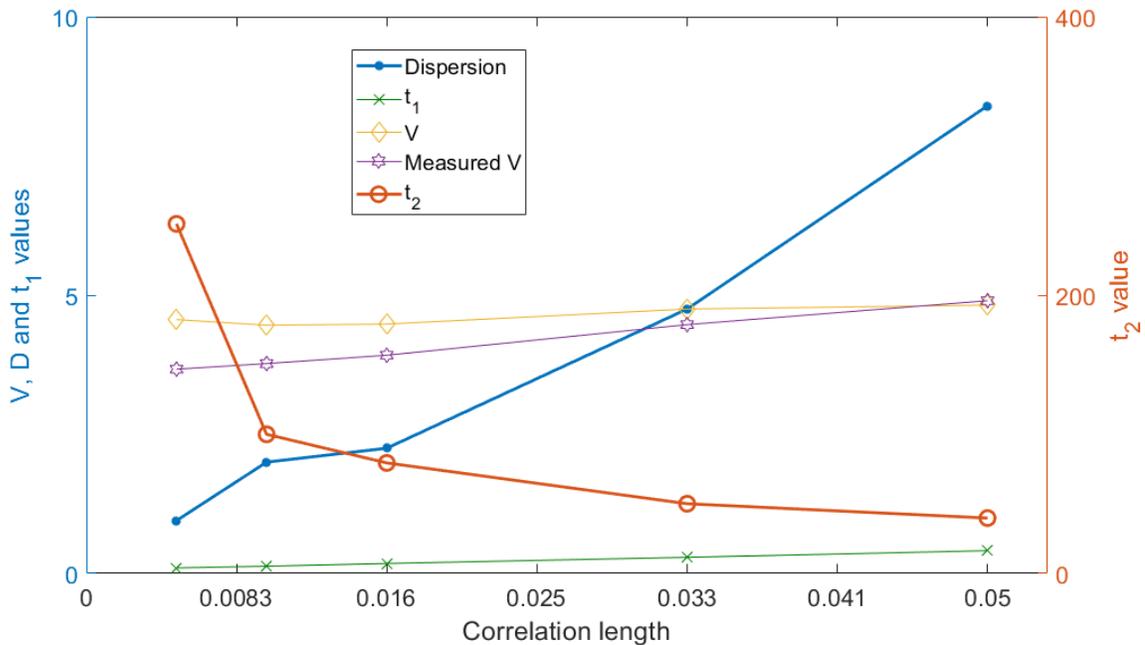

Figure 6. Parameters of the CTRW-TPL fits to ensemble BTCs with correlation lengths of $5\times10^{-3}$, $1\times10^{-2}$, $1.6\times10^{-2}$, $3.3\times10^{-2}$, and $5\times10^{-2}$. While $t_2$ decreases at an exponential rate, the other parameters increase with a linear rate and with lower values.